# Logic Circuits Using Solution-Processed Single-Walled Carbon Nanotube Transistors

Ryo Nouchi[a)], Haruo Tomita, Akio Ogura and Masashi Shiraishi

*Division of Materials Physics, Graduate School of Engineering Science, Osaka University, Machikaneyama-cho 1-3, Toyonaka 560-8531, Japan*

Hiromichi Kataura

*Nanotechnology Research Institute (NRI), National Institute of Advanced Industrial Science and Technology (AIST), Central 4, Higashi 1-1-1, Tsukuba, Ibaraki 305-8562, Japan*

This letter reports on the realization of logic circuits employing solution-processed networks of single-walled carbon nanotubes.  We constructed basic logic gates (inverter, NAND and NOR) with n- and p-type field-effect transistors fabricated by solution-based chemical doping.  Complementary metal-oxide-semiconductor inverters exhibited voltage gains of up to 20, which illustrates the great potential of carbon nanotube networks for printable flexible electronics.

[a)] Author to whom correspondence should be addressed; electronic mail: nouchi@mp.es.osaka-u.ac.jp

Flexible electronic devices constructed on plastic substrates have being attracting considerable attention since they have many applications, including electronic paper and wearable computing systems, which are not achievable using conventional silicon-based electronics. Organic molecules,[1] inorganic materials,[2] and single-walled carbon nanotubes (SWNTs)[3,4] have been investigated as the fundamental components for this emerging field, exploiting their capabilities for low-temperature processing. Out of these candidates, SWNTs show robust flexibility and are thus very promising for realizing flexible electronics.[5] Field-effect transistors (FETs) using SWNTs as a channel are compatible with flexible plastic substrates in the form of thin-film transistors (TFTs) with a solution-processed random network of SWNTs.[6,7] SWNT TFTs simultaneously exhibit high field-effect mobilities (3.6 $cm^2$ $V^{-1}$ $s^{-1}$) and large on/off current ratios ($10^4$).[7] Earlier studies reported mobilities of higher than 10 $cm^2$ $V^{-1}$ $s^{-1}$; however, their on/off ratios were restricted to $10^2$ at best due to a trade-off relationship between the mobility and the on/off current ratio[6]. The mobility of 3.6 $cm^2$ $V^{-1}$ $s^{-1}$ is higher than that of any organic TFT with solution-processed channels, and is comparable to the highest mobility reported for organic TFTs with vacuum-evaporated channels. Recently, a transparent flexible TFT using an SWNT network was demonstrated and it was found to be much stronger against bending than a transparent flexible TFT with an inorganic oxide channel.[3] Therefore, SWNT TFTs are considered to be a major candidate for future solution-processable electronics, i.e., *printable* flexible electronics.

In this study, we fabricated n- and p-type SWNT TFTs employing solution processes and succeeded in constructing complementary metal-oxide-semiconductor (CMOS) logic circuits. The simplest logic circuit, an inverter, exhibited a voltage gain of up to 20. This is the highest gain for inverters with solution-processed TFTs. This is a fundamental step toward realizing printable flexible electronics.

SWNTs were synthesized using a laser ablation method and then purified using a conventional method.[8] These SWNTs typically had diameters of approximately 1.4 nm and a length of several micrometers. After purification, the SWNTs were ultrasonicated in dimethylformamide solution for 5 hours to dissolve the SWNT bundles. They were then centrifuged at 12300 rpm for 15 min to select well-dispersed, narrow bundles of the SWNTs. The centrifuged solution was decanted and dispersed on a heavily doped Si substrate with a 200-nm-thick thermal oxide layer on top of it. The Si substrate was used as a back gate electrode. Prior to dispersion, source and drain electrodes of 70-nm-thick Au films with a 10-nm-thick Cr adhesion layer were patterned on the substrate by electron beam lithography and lift-off techniques. The channel length and width were designed to be 5 and 100 μm, respectively. A schematic of the transistor structure is shown in Fig. 1. Since metallic-SWNTs (m-SWNTs) were formed in addition to semiconducting SWNTs as a channel of the SWNT TFTs, a low on/off ratio posed a serious problem, owing to the lack of electric field response of m-SWNTs. To remove the m-SWNTs, we introduced an electrical breakdown.[9,10] The detailed breakdown procedure is described in a previous paper.[7] Hereafter, we discuss the device characteristics achieved after the breakdown procedure. Device characterization was performed in a low vacuum at room temperature.

Since as-dispersed SWNTs exhibit intrinsically ambipolar FET characteristics, in order to construct CMOS logic circuits it is necessary to fabricate unipolar n- and p-type FETs. The CMOS logic has low power consumption compared with other logic families, such as the resistor-transistor logic. It has been shown that stable n- and p-type SWNT FETs can be obtained by chemical doping of polyethyleneimine (PEI)[11,12] and tetracyano-$p$-quinodimethane (TCNQ)[12,13], respectively. We employed chemical doping to control the polarity of SWNT TFTs, using methanol and $CS_2$ as solvents for PEI and TCNQ doping, respectively. This solution-based doping method should also be compatible with printable electronics. Figure 2(a) shows the transfer characteristics of the fabricated PEI- and TCNQ-doped SWNT TFTs; these devices clearly display n- and p-type FET characteristics, respectively.

The ability to control the polarity of SWNT TFTs enables us to construct complementary logic circuits. Figure 2(b) shows the output characteristics of the simplest logic gate, an inverter (NOT), utilizing n- and p-type FETs produced by the chemical doping method. This device displays a voltage gain of approximately 20, which is the highest value reported for CMOS inverters with solution-processed channels. The SWNT TFTs that compose the inverter have a field-effect mobility of about $10^{-2}$ cm$^2$ V$^{-1}$ s$^{-1}$ and an on/off current ratio of $10^4$-$10^5$. Although this mobility is moderate for a solution-processed TFT, it is very low for a solution-processed SWNT TFT[7], indicating that the gain can be further improved by optimizing the fabrication parameters. This result clearly illustrates the great potential of the SWNT TFT as a basic component in flexible electronics.

We further investigated the potential of SWNT TFTs by constructing more complex logic circuits. Figure 3 shows the output characteristics of other basic logic gates (NAND and NOR). These complementary logics are constructed with two n-type and two p-type SWNT TFTs. Proper device operation was confirmed.

The ultimate test for device performance is to demonstrate a ring oscillator in which an odd number of inverters are connected in series. Such a circuit has no stable solution and the output voltage of each inverter stage oscillates as a function of time. The oscillation frequency of the circuit is determined by the propagation delay in voltage switching between successive inverter stages. This delay depends on the $RC$ constant of the circuit. Thus, the parasitic capacitance governs the oscillation frequency. The n- and p-type SWNT TFTs were fabricated on different substrates in this study, due to the inability to perform local chemical doping on the same substrate. A ring oscillator was not constructed in this study since the oscillation frequency is inevitably affected by the parasitic capacitance of the off-chip wiring between the n- and p-type TFTs.

The most promising method for achieving local chemical doping is inkjet printing. This method will enable us to fabricate FETs of both polarities on the same substrate by local deposition of donor or acceptor molecules, enabling on-chip wiring to be used. By varying the amount of dopants in chemical doping, it is possible to control the threshold voltage of SWNT FETs.[14] Therefore, local inkjet doping has the potential to tune the threshold voltage of an SWNT TFT by varying the donor/acceptor molecule concentration in the ink or varying the number of printings. By fine-tuning the threshold voltage it will be possible to improve the performance of CMOS logics. Furthermore, the printing method is applicable to the SWNT-network channel itself. The realization of inkjet-processed transparent flexible SWNT TFTs will be described elsewhere.[15]

In summary, we reported on the realization of CMOS logic circuits constructed using solution-processed SWNT TFTs. The fabricated CMOS inverters showed voltage gains of up to 20, which is the highest value for inverters with solution-processed TFTs. In this study, n- and p-type TFTs were fabricated on different substrates by PEI and TCNQ doping, respectively. By employing an inkjet printing method, it should be possible to integrate SWNT TFTs of both polarities onto the same substrate. This study demonstrates that high-performance CMOS logics can be constructed from solution-processed SWNT TFTs. This type of FET is highly compatible with low-temperature printing techniques, and SWNT TFTs should be a major candidate in the emerging field of printable flexible electronics.

FIGURE CAPTIONS

FIG. 1. (Color online) Schematic diagram of the structure of the SWNT TFTs.

FIG. 2. (Color online) (a) Transfer characteristics of PEI- and TCNQ-doped SWNT TFTs.

(b) Output characteristics of an inverter constructed with PEI-doped (n-type) and TCNQ-

doped (p-type) SWNT TFTs.  This device has a voltage gain of around 20.

FIG. 3. (Color online) Output characteristics of (a) NAND and (b) NOR gates.  These

CMOS logic circuits are constructed from two n-type and two p-type SWNT TFTs.

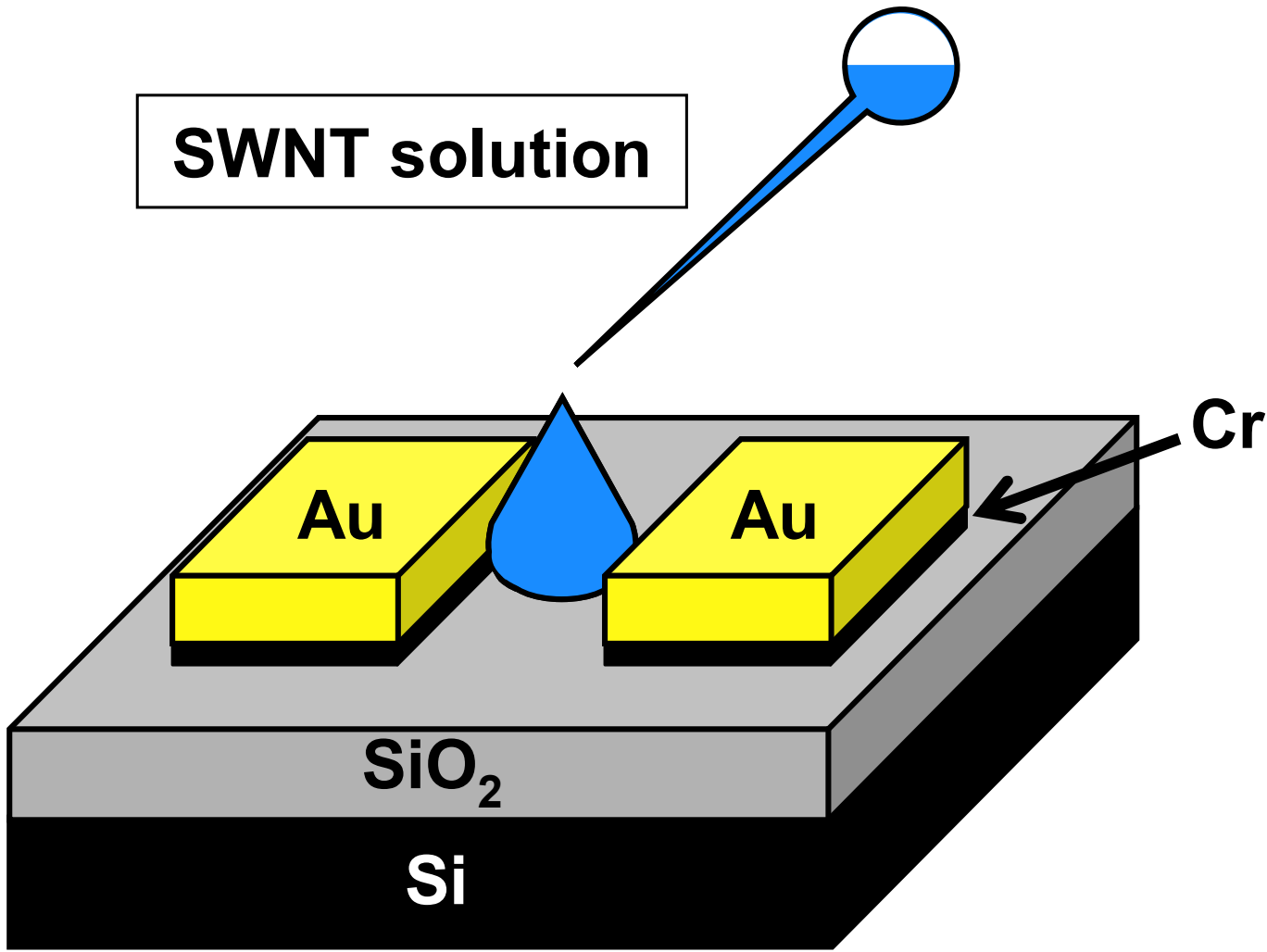

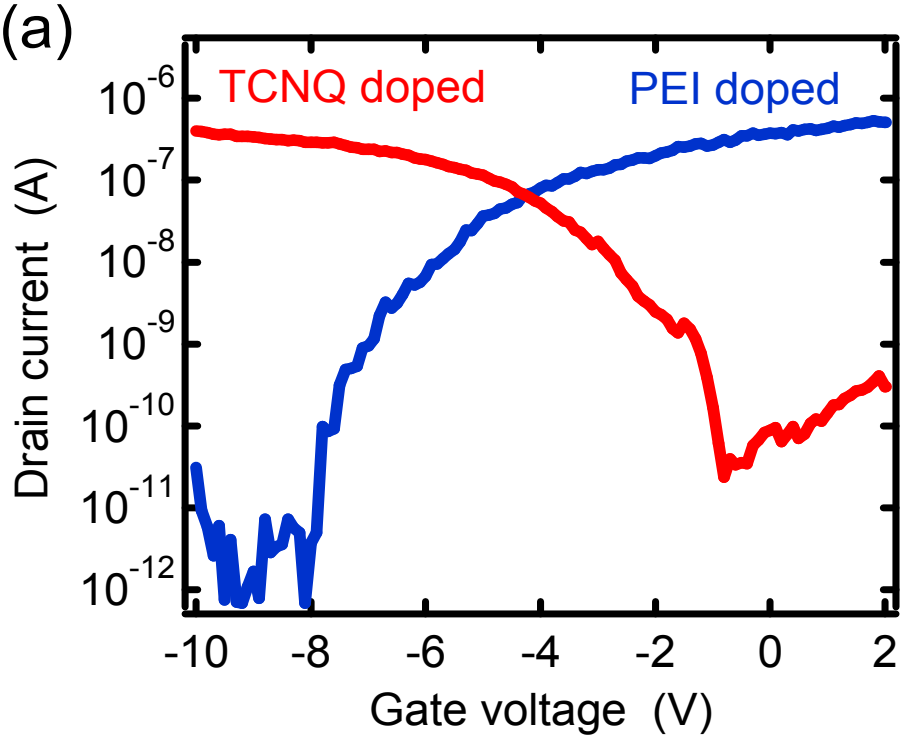

(a)

TCNQ doped     PEI doped

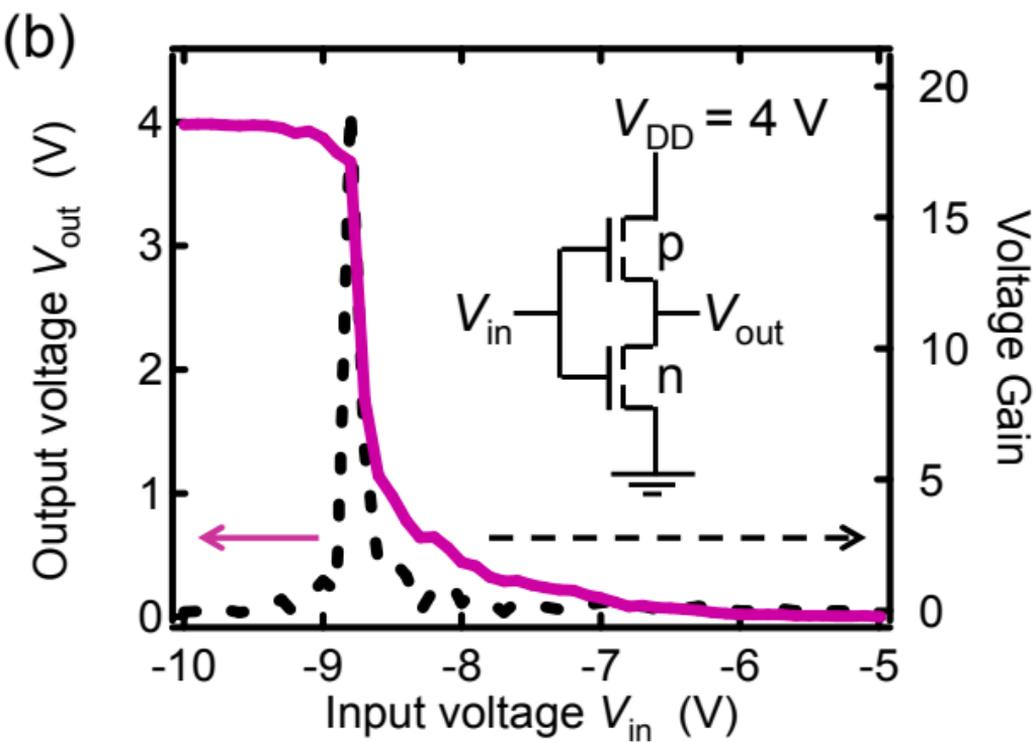

(b)

$V_{DD} = 4$ V

p

$V_{in}$     $V_{out}$

n

# (a) NAND

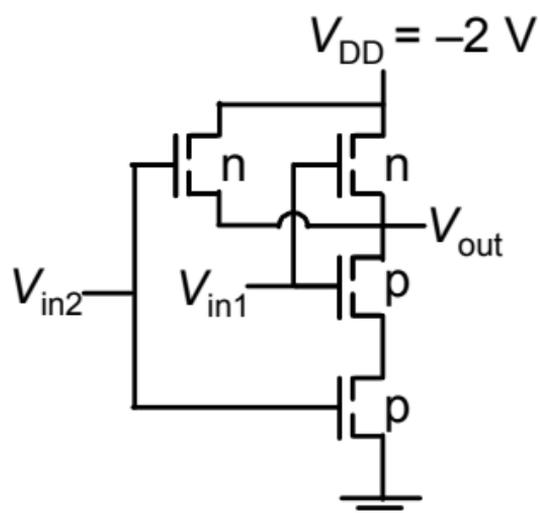

| in1 | in2 | out |
|-----|-----|-----|
| 1 | 1 | 0 |
| 1 | 0 | 1 |
| 0 | 1 | 1 |
| 0 | 0 | 1 |

$V_{DD} = -2$ V

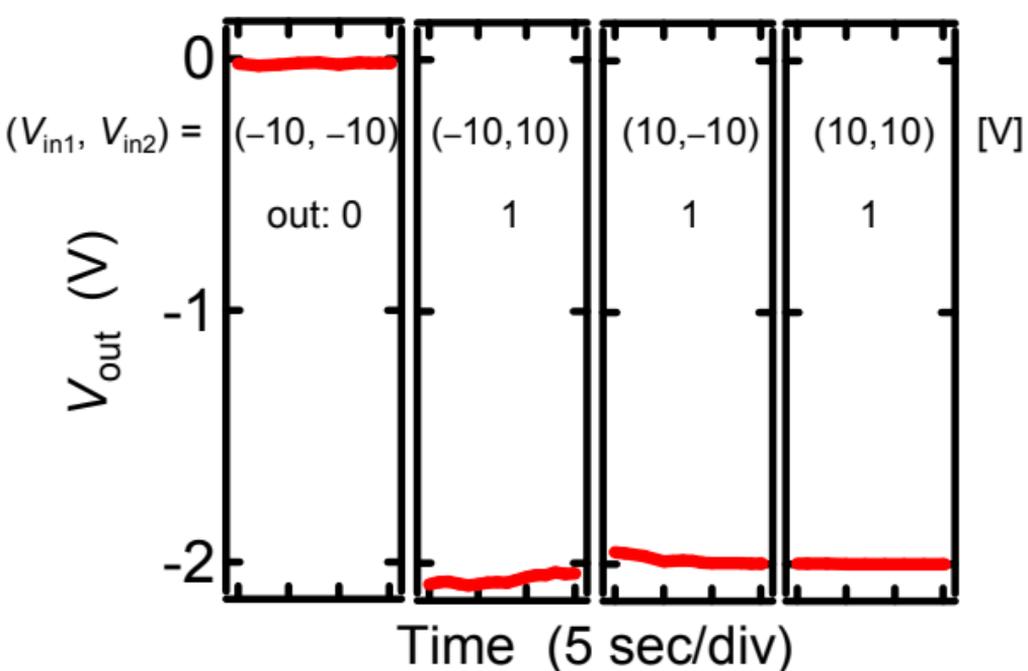

$(V_{in1}, V_{in2}) =$ (−10, −10)  (−10,10)  (10,−10)  (10,10)  [V]

out: 0   1   1   1

$V_{out}$ (V)

Time  (5 sec/div)

# (b) NOR

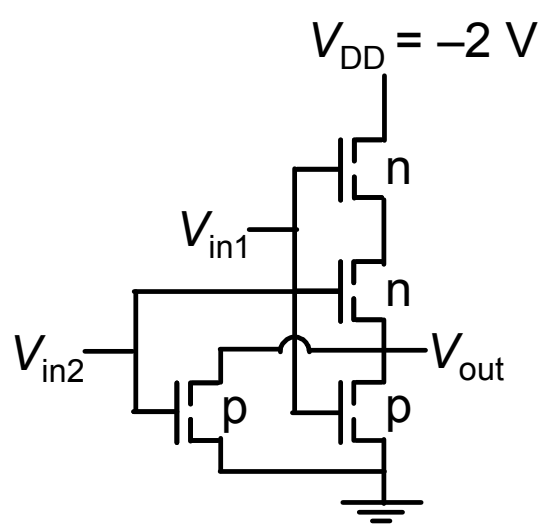

| in1 | in2 | out |
|-----|-----|-----|
| 1 | 1 | 0 |
| 1 | 0 | 0 |
| 0 | 1 | 0 |
| 0 | 0 | 1 |

$V_{DD} = -2$ V

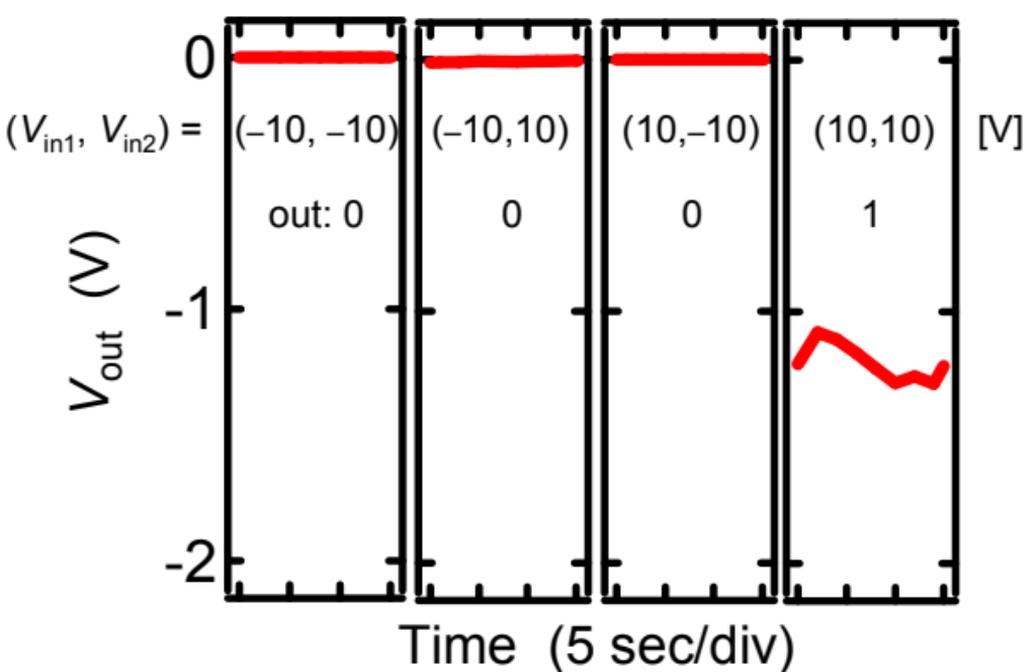

$(V_{in1}, V_{in2}) =$ (−10, −10)  (−10,10)  (10,−10)  (10,10)  [V]

out: 0   0   0   1

$V_{out}$ (V)

Time  (5 sec/div)